\documentclass[12pt,preprint]{aastex}
\usepackage{float}
\usepackage{arydshln}
\newcommand{\Teff}{\mbox{$T_{\rm eff}$}}

\newcommand{\Lya}{Lyman~$\alpha$}
\newcommand{\lya}{Lyman~$\alpha$}
\newcommand{\hst}{{\sl HST}}
\newcommand{\stis}{{\sl STIS}}

\newcommand{\galex}{{\sl GALEX}}
\newcommand{\COS}{{\sl COS}}

\slugcomment{Accepted to ApJL}

\shorttitle{PREDICTING \lya\ and Mg II FLUXES FROM GALEX UV PHOTOMETRY
}
\shortauthors{Shkolnik et al.}

\begin{document}


\title{Predicting \lya\ and Mg II Fluxes from K and M Dwarfs Using \galex\ Ultraviolet Photometry
\altaffilmark{1}\\}


\author{Evgenya~L.~Shkolnik}
\affil{Lowell Observatory, 1400 West Mars Hill Road, Flagstaff, AZ, 86001, USA}
\email{shkolnik@lowell.edu}

\author{Kristina~A.~Rolph}
\affil{Lowell Observatory, 1400 West Mars Hill Road, Flagstaff, AZ, 86001, USA}
\affil{Franklin and Marshall College, Lancaster, PA 17604, USA}
\email{kristina.rolph@fandm.edu}

\author{Sarah Peacock}
\affil{Department of Planetary Sciences and Lunar and Planetary Laboratory\\
University of Arizona, Tucson AZ, 85721, USA}
\email{speacock@lpl.arizona.edu}

\and

\author{Travis S. Barman}
\affil{Department of Planetary Sciences and Lunar and Planetary Laboratory\\
University of Arizona, Tucson AZ, 85721, USA}
\email{barman@lpl.arizona.edu}

\altaffiltext{1}{Based on observations made with the NASA \emph{Galaxy Evolution Explorer}.
\galex\ was operated for NASA by the California Institute of Technology under NASA contract NAS5-98034.}

\begin{abstract}

A star's UV emission can greatly affect the atmospheric chemistry and physical properties of closely orbiting planets with the potential for severe mass loss. 
In particular, the \lya\ emission line at 1216\AA, which dominates the far-ultraviolet spectrum, is a major source of photodissociation of important atmospheric molecules  such as water and methane. 
The intrinsic flux of \lya, however, cannot be directly measured due to the absorption of neutral hydrogen in the interstellar medium and contamination by geocoronal emission. 
To date, reconstruction of the intrinsic \lya\ line based on \emph{Hubble Space Telescope} spectra has been accomplished for 46 FGKM nearby stars, 28 of which have also been observed by the \emph{Galaxy Evolution Explorer} (\galex). Our investigation provides a correlation between published intrinsic \lya\  and \galex\ far- and near-ultraviolet chromospheric fluxes for K and M stars.  
The negative correlations between the ratio of the \lya\ to the \galex\ fluxes reveal how the \emph{relative} strength of \lya\ compared to the broadband fluxes weakens as the FUV and NUV excess flux increase.
We also correlate \galex\ fluxes with the strong near-ultraviolet Mg~II~h+k spectral emission lines formed at lower chromospheric temperatures than \lya.  The reported correlations provide estimates of intrinsic \lya\ and Mg~II fluxes for the thousands of K and M stars in the archived \galex\ all-sky surveys. These will constrain new stellar upper-atmosphere models for cool stars and provide realistic inputs to models describing exoplanetary photochemistry and atmospheric evolution in the absence of ultraviolet spectroscopy.

\end{abstract}


%%%%%%%%%%%%%%%%%%%%%%%%%%%%%%%%%%%%%%%%%%%%%%%%%%%%%%%%%%%%%%%%%%%%%%%%%%%%%%%%%%%%%%%%%%%%%%%%%%%%%%%%%%%%%%%%%%%%%%%%%%%%%%

\section{Introduction}\label{intro}

Radial velocity, transit, and imaging methods have enabled the discovery of a variety of exoplanets ranging from hot super Earths to cold Jupiters, including confirmed terrestrial exoplanets within the habitable zone (HZ; \citealt{kast93}) around low-mass stars. For giant and terrestrial planets alike, incident stellar emission at short wavelengths ($\lambda <$ 3000\AA) affects the chemistry and evolution of an exoplanet's atmosphere. In particular, it is the high energy of the far-ultraviolet (FUV) radiation that controls the processes of molecular photodissociation and photoionization in the planetary atmosphere (\citealt{hu12}), breaking apart important molecules such as water, methane and carbon dioxide, and potentially leading to significant mass loss. The possibility of complete evaporation of the planet atmosphere due to high UV (and corresponding particle) flux may explain the high fraction of hot dense planets around cool stars in the current exoplanet population (e.g.~\citealt{wu13,lamm07}). In the case of HZ planets, the incident UV stellar flux may also destroy biosignatures with which we hope to detect life, and/or produce false-positive biosignatures  in the form of abiotic oxygen and ozone \citep{tian14}.

Much of the stellar UV flux is emitted from the upper-atmospheres of FGKM stars, namely the chromosphere, transition region and corona, and  measurements at these short wavelengths require space-borne observatories and stars that are bright and not too distant from Earth. The FUV spectrum is dominated by  \lya, the resonance line of hydrogen at  1216 \AA,  which is emitted from the upper-chromosphere and lower-transition region temperatures of roughly 8000  to 30000 K \citep{font88}. \cite{lins13} recently highlighted the potential importance of the \lya\ emission line to the atmospheres of close-in planets.  For example, the Sun's \lya\  contributes  $\sim$20\% of the total flux between 1 to 1700 \AA\ \citep{riba05}, and is shown to be responsible for more than 50\% of the photodissociation rate of the Earth's H$_2$O and CH$_4$.
\lya's fractional flux increases for cooler stars as the photospheric contribution decreases (Fig.~\ref{Teff_phot}, right) and can reach up to $\sim$70\% of the total FUV flux for late M dwarfs \citep{fran13}.

In addition to the observational difficulties, predicting stellar UV flux from low-mass stars is also challenging since current models were not designed to calculate emission from the low-density regions of stellar upper-atmospheres. As a result, the UV flux received by an exoplanet is often underestimated in photochemical models \citep{kopp12,migu14a,tian14}. Recently, \cite{migu14b} demonstrated the significant impact of a star's \lya\ on the photochemistry of the atmospheres of  mini-Neptunes at low pressures  using the GJ 436 planetary system, one of the few M dwarfs for which \lya\ data exist \citep{fran13}.  
	
	The total \lya\ emission is impossible to measure directly from stars due to scattering of the majority of the photons by the neutral hydrogen of the interstellar medium. Even for the nearest stars it is difficult to directly determine the \lya\ flux without using reconstruction techniques applied to high-resolution spectra  currently available only with the \emph{Hubble Space Telescope}'s (\hst) \emph{Space Telescope Imaging Spectrograph} (\stis) and \emph{Cosmic Origins Spectrograph} (\COS). %\citep{wood05,fran12,fran13}.
 Complicating the measurement further, one also has to take into account Earth's own geocoronal \lya\ emission. 
	
	Using  reconstructions from \cite{wood05} and \cite{fran12,fran13},  \cite{lins13} produced correlations of \lya\ line fluxes from a sample 46 FGKM stars with other spectral emission lines formed in the chromosphere,  which were also observed with either \stis\ or \COS. 
	Although the Mg II h+k emission lines at 2802.7 and 2795.5 \AA\ also required a correction for interstellar absorption at their cores \citep{wood05,lins13}, they provided the most promising results possibly due to being least affected by metallicity differences.  Even with large uncertainties in the \lya\ reconstructed fluxes,  $\approx$20\% for the FGK stars \citep{wood05} and $\approx$30\% for the M stars \citep{fran12,fran13}, the search for correlations between different activity diagnostics and \lya\ is extremely valuable when no other UV spectral data are available. 

In view of the severe challenges of measuring or predicting \lya\ and Mg~II for many stars, we used \emph{The Galaxy Evolution Explorer}'s (\galex) all-sky photometric surveys to search for correlations between FUV and NUV photometry and the intrinsic \lya\ and Mg~II fluxes compiled by \cite{lins13}. In the absence of UV spectroscopy, such correlations will allow for estimates of the intrinsic \lya\ fluxes from thousands of stars within a few hundred parsecs in the \galex\ archive with which to study stellar activity, provide empirical constraints for new cool-star upper-atmosphere models (S.~Peacock et al., in preparation) and estimate the incident high-energy radiation affecting the photochemisty and evolution of  planetary atmospheres.

%%%%%%%%%%%%%%%%%%%%%%%%%%%%%%%%%%%%%%%%%%%%%%%%%%%%%%%%%%%%%%%%%%%%%%%%%%%%%%%%%%%%%%%%%%%%%%%%%%%%%%%%%%%%%%%%%%%%%%%%%%%%%%

	\section{\galex\ NUV and FUV Photometry}

	The \galex\ satellite was launched on 2003 April 28 and observed the UV sky until its mission completion on 2013 June 28. 
	\galex\ imaged approximately 3/4 of the sky simultaneously in two UV bands: FUV 1350--1750 \AA\ and NUV 1750--2750 \AA, excluding the powerful and complicated \lya\ and nominally sensitive to the Mg II h+k chromospheric lines.  The average FWHM of the PSFs are 6.5\arcsec\ and 5\arcsec\ in the FUV and NUV, respectively, across a 1.25$^{\circ}$ field of view. The full description of the instrumental performance is presented by \cite{morr05}.\footnote{One can query the \galex\ archive through either CasJobs (http://mastweb.stsci.edu/gcasjobs/) or the web tool GalexView (http://galex.stsci.edu/galexview/).} With the failure of the FUV detector in 2009 May, subsequent observations only provided NUV imaging. The fluxes and magnitudes averaged over the entire exposure were produced by the standard \galex\ Data Analysis Pipeline (ver.~4.0) operated at the Caltech Science Operations Center \citep{morr07}. The current database contains 214,449,551 source measurements (Bianchi 2013) recorded by the All-sky, Medium and Deep Imaging Surveys and many guest investigator programs.  The data products are archived at the Barbara A. Mikulski Archive for Space Telescopes (MAST).
	
For this study we used the ``aper\_7'' aperture for the UV photometry, which has a radius of 17.3\arcsec. This large aperture requires the least aperture correction (0.04 mags in both the in NUV and FUV bandpasses)\footnote{See Table 1 of http://www.galex.caltech.edu/researcher/techdoc-ch5.html. Note \cite{morr07} quote a required aperture correction of 0.07 mags.  Either way the effect is very small compared to the uncertainties and the large differences in flux between targets.} and accounts for all possible point spread functions, even those elongated near the edges of the images. We excluded detections beyond 0.59$^\circ$ from the center of the  image to avoid the worst edge effects.

 For F and G stars, the flux in the \galex\ bandpasses includes a significant fraction of continuum (i.e.~photospheric) emission (Fig.~\ref{Teff_phot}, right; \citealt{smit10}) with the remaining flux provided by strong emission lines (C IV, C II, Si IV, He
II) originating from the stellar upper-atmosphere. K and M stars have FUV and NUV fluxes strongly dominated by these stellar emission lines making \galex\ an excellent tool with which to study stellar activity in lower-mass stars (e.g.~\citealt{robi05,wels06,paga09a,find10,shko11a,rodr11,shko13}).

Twenty-eight of the stars with reconstructed \lya\ fluxes compiled by \cite{lins13} were observed by \galex.
 These are summarized in Table~\ref{table_targets}. 
\galex\ observations move into the non-linear regime after 34 counts~s$^{-1}$ in the FUV and 108 counts~s$^{-1}$ in the NUV \citep{morr07}. This required us to omit all the bright stars with effective temperatures \Teff\ $>$ 4900 K from the NUV analysis and one nearby K star from the FUV analysis. One M star has an FUV flux below the detection threshold and we calculate its 1-sigma upper limit using Figure 4 of \cite{shko14}.  This left eight stars with NUV detections and nineteen stars with FUV detections for analysis.

%%%%%%%%%%%%%%%%%%%%%%%%%%%%%%%%%%%%%%%%%%%%%%%%%%%%%%%%%%%%%%%%%%%%%%%%%%%%
\section{Analysis}\label{results}

\cite{lins13} compiled the spectral types (SpTs), distances, and reconstructed \lya\ and Mg~II fluxes scaled to 1 AU of 46 stars. Table~\ref{table_targets} lists the 28 of these stars observed by \galex. We scaled the fluxes to the stellar surface using radii derived from the \cite{bara98} models with published stellar ages from the literature and \Teff\ from \cite{krau07} for the given SpT. 

\begin{figure}[tbp]
	\plottwo{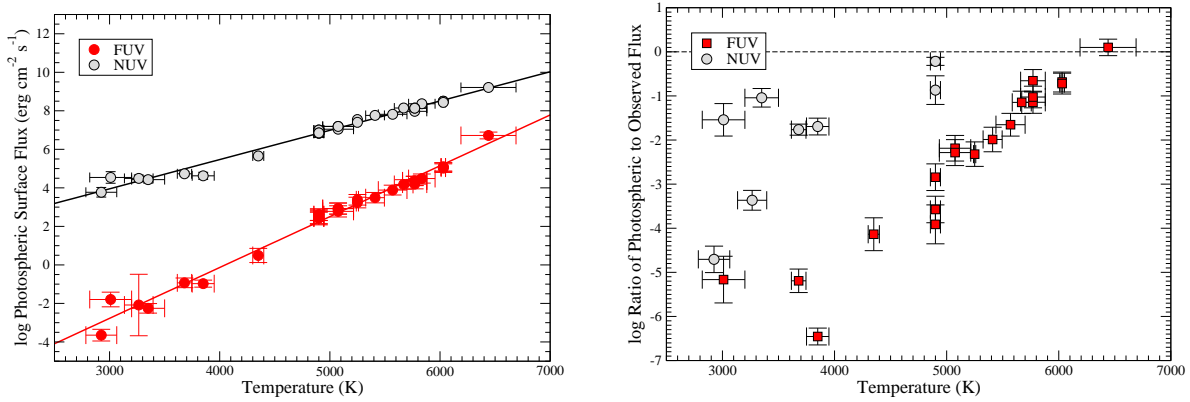}{teff_ratio.eps}
	\caption{\emph{Left:} FUV and NUV photospheric surface fluxes as a function of effective temperature for our sample. The M stars (2900 -- 4000 K) use main-sequence photospheric fluxes from the Phoenix models while the FGK stars (4000 -- 6500 K) use the Kurucz-based values from \cite{find11} for which empirically-deduced \Teff\ offsets were applied. The coefficients of the regression analysis are listed in Table~\ref{table_coef}. \emph{ Right:} Ratio of the photospheric flux to the total observed surface flux for \galex\ FUV and NUV bandpasses as a function of temperature.\label{Teff_phot}}
\end{figure}

As mentioned above, the \galex\ bandpasses consist of both photospheric and chromospheric emission. We determined photospheric emission from stellar atmosphere models, which by design exclude chromospheric emission, to isolate the excess flux originating solely from upper-atmospheric activity.
We adopt the semi-empirical results from  \cite{find11} who calculated \galex\ FUV and NUV magnitudes of the photospheric fluxes for B8 to K5 stars using the solar-metallicity Kurucz models and applied \Teff\ corrections to fit empirical color-color plots. (See their Table 1 and Figure 5.) 
For stars cooler than K5, we use the Phoenix model atmospheres (\citealt{haus97,shor05}) convolved
with the relevant NUV and FUV normalized transmission
curves. (See Section~3 of \citealt{shko14}).   
Fig.~\ref{Teff_phot} (left) plots the surface photospheric flux for our sample as a function of \Teff:  the \cite{find11} fluxes for stars hotter than 4000 K and  Phoenix model fluxes for cooler stars. The tight correlation between the two photospheric flux sources demonstrates their consistency.

\begin{figure}[tbp]
	\includegraphics[width=5.5in,angle=90]{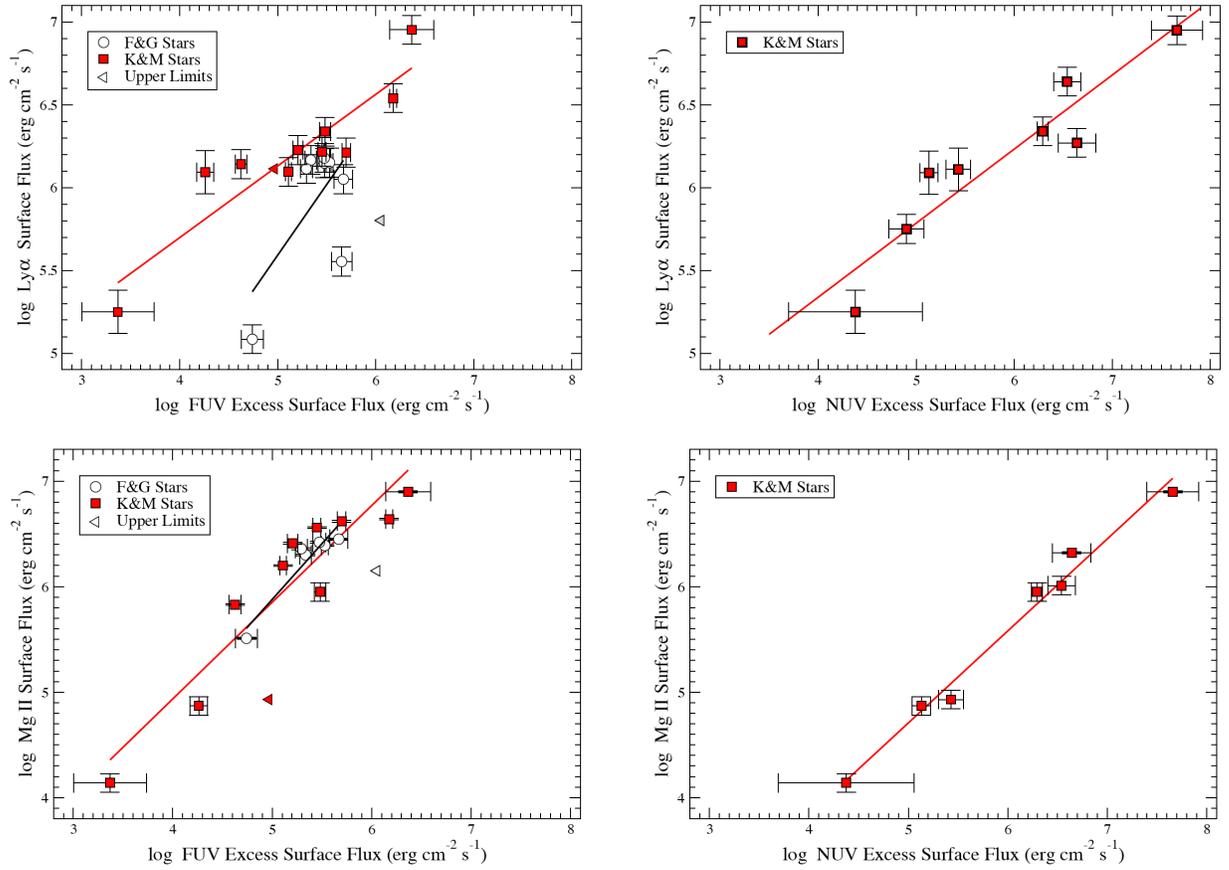}
		\caption{FUV (left column) and NUV (right column) excess surface flux plotted against \lya\  (top row) and Mg II (bottom row) surface flux. \lya\ and Mg II fluxes are taken from \cite{lins13} and scaled to the stellar surface.\label{uv_lya}}
	\end{figure}

In both the NUV and FUV detections of our sample, 80\% of the targets have photospheric contributions of $<$10\% of the observed flux. Only one hot star at \Teff=6800 K appears to have nearly no chromospheric emission. Fig.~\ref{Teff_phot} (right) plots the ratio of photospheric flux to the observed  FUV and NUV surface fluxes as a function of \Teff\ for our sample. All but one of the K and M stars have negligible photospheric contributions. The observed and excess NUV and FUV fluxes for the sample are listed in Table~\ref{table_targets}.   Although in most cases the expected photospheric flux is relatively low, we subtract it from the \galex\ observations such that correlations can be made using only upper-atmospheric emission with the \lya\ and Mg~II chromospheric line (Fig.~\ref{uv_lya}). Within uncertainties, correlations for the K and M stars are unchanged by this step. (See Table~\ref{table_coef}.)

We separated the stars into two SpT bins: F and G stars and K and M stars for several reasons:  (1) The fraction of the photosphere in the hotter stars is much higher than in the cooler stars for which it is effectively negligible. Thus, any uncertainties in the models used will not disrupt the results for the K and M stars, as it might for the hotter stars. (2) Low mass stars tend to have higher flare activity levels, so the non-contemporaneous data sets used in this study may induce added scatter for K and M stars.\footnote{\hst\ observations of the old M dwarf GJ 876 ($\sim$3 Gyr; \citealt{corr10}) show flaring in upper-atmospheric emission lines with flux levels increasing by at least a factor of 10 during a 5800-second observation \citep{fran12}.} (3) And, with K and M stars being the most favorable for followup studies of HZ planets (e.g.~\citealt{tart07,scal07,hell14}),
a separate correlation for these stars is appropriate for providing \Lya\ and Mg~II flux estimates with which to study planetary atmospheres.

Fig.~\ref{uv_lya} (left) shows the FUV excess  to \lya\ surface flux correlation for the 11 K and M stars with a correlation coefficient $R$=0.91, excluding the one upper limit. The F and G correlation has eight stars with a weaker correlation coefficient R=0.63 and low statistical significance. 
Fig.~\ref{uv_lya} (right) shows the correlation for the \lya\ surface flux against the NUV excess surface flux with a strong correlation of $R$=0.94 for eight K and M stars.\footnote{Note that the strongest UV emitting M star in the sample is the young Speedy Mic (40 Myr; \citealt{krau14}) and the weakest emitting is the old GJ 876. \cite{clai12} have shown that stellar high-energy flux decreases with age for a sample of Sun-like stars with steepening decline at shorter wavelengths. The same is true for M dwarfs (\citealt{prei05,shko14}).} Due to the brightness of the F and G stars, no NUV observations were reliable due to the non-linearity of the \galex\ data.

\begin{figure}[tbp]
		\plottwo{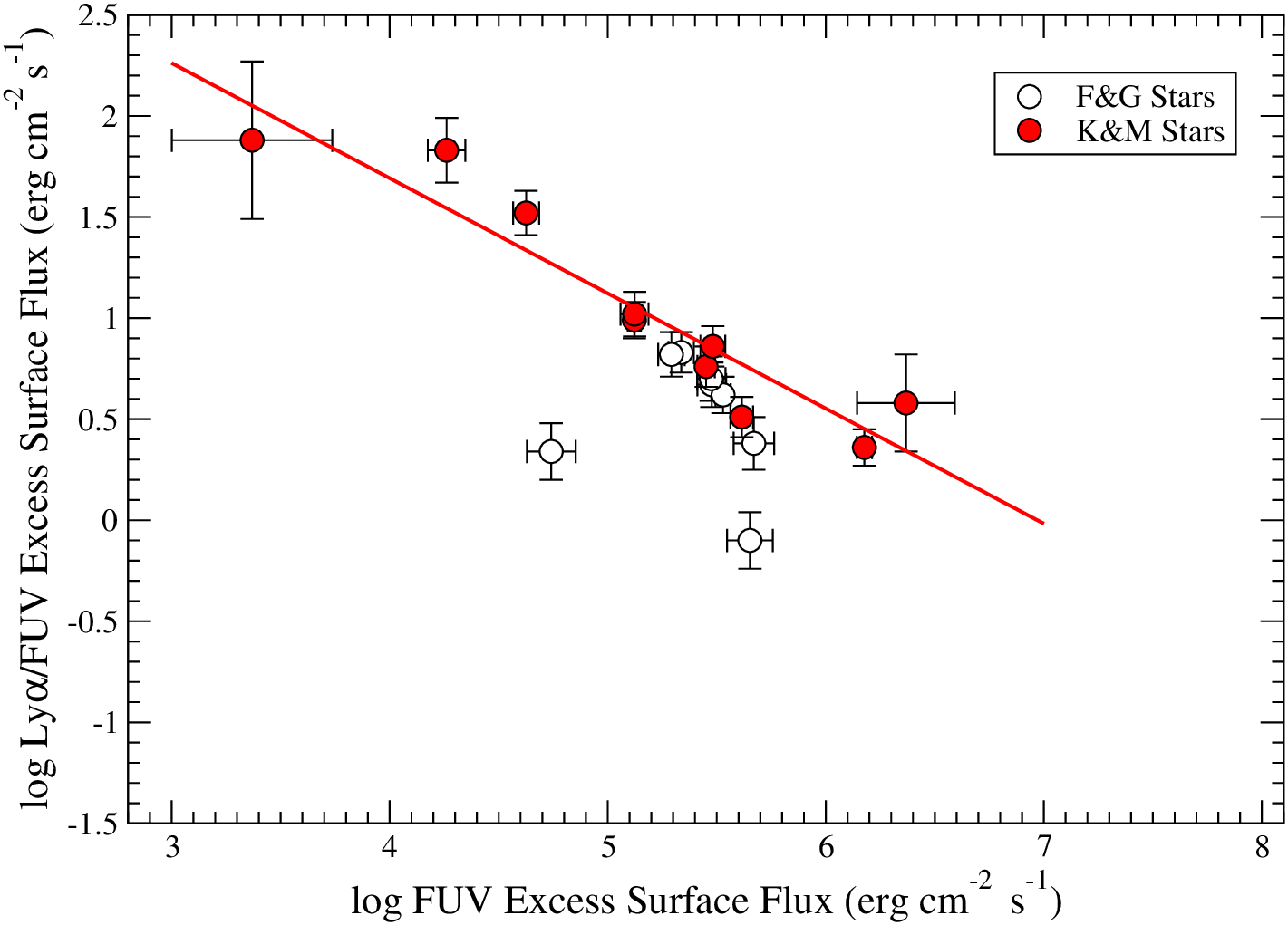}{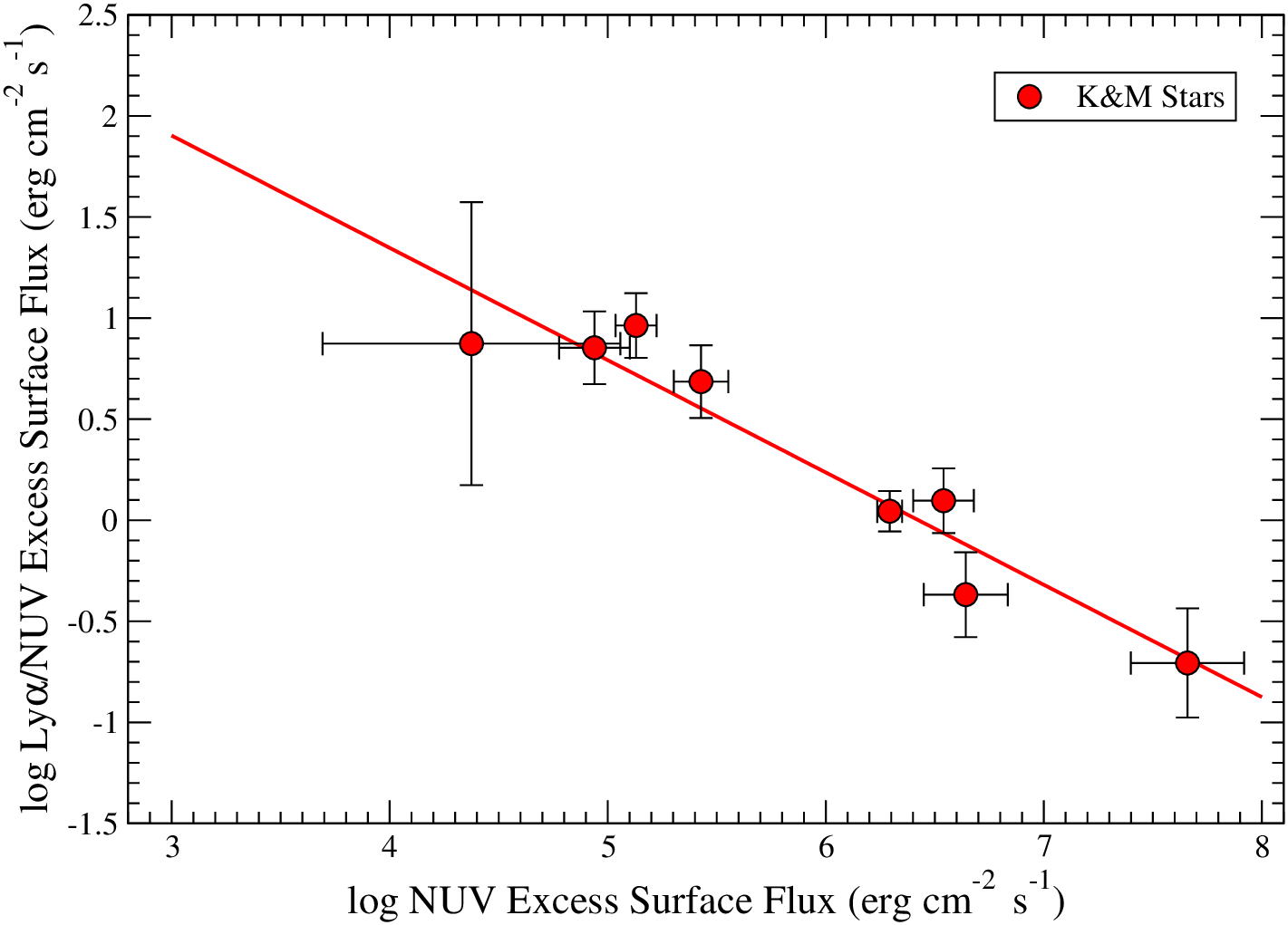}
		\caption{Ratio of \lya\ to FUV (left) and NUV (right) surface flux plotted against the FUV and NUV excess surface flux. \label{uv_lya_ratio}}
	\end{figure}

\cite{lins13} pointed out that of all the emission lines they studied, Mg II may provide the most accurate \lya\ predictions as metallicity effects appear the smallest. Our analysis also exhibits highly significant correlations between \galex\ fluxes and the Mg~II lines for the K and M stars (Fig.~\ref{uv_lya}, bottom).
Fig.~\ref{uv_lya_ratio} shows the ratio of \lya\ to the \galex\ flux as a function of FUV and NUV flux. These correlations reveal how the \emph{relative} strength of the \lya\ compared to the broadband fluxes weakens as the FUV and NUV excess flux increases.  All of the regression fits to the correlations are summarized in Table~\ref{table_coef}.

%%%%%%%%%%%%%%%%%%%%%%%%%%%%%%%%%%%%%%%%%%%%%%%%%%%%%%%%%%%%%%%%%%%%%%%%%%%%%%%%%%%%%%%%%%%%%%%%%%%%%%%%%%%%%%%%%%%%%%%%%%%%%%

\section{Summary}\label{summary}

\lya\ is the strongest stellar UV emission line and provides the most important source of radiative losses in a star's upper-chromosphere and lower-transition region. As such, it may also be the greatest source of molecular photodissociation in the atmosphere of an orbiting exoplanet. Measuring a star's intrinsic \lya\ flux requires space-based high-resolution FUV spectroscopy plus intricate reconstruction techniques to correct for interstellar absorption and geocoronal emission.   In order to gain access to the \lya\ emission of many more stars than is currently possible to observe (i.e.~with \hst), we sought to find correlations between literature values of reconstructed \lya\ emission and broadband stellar upper-atmospheric emission measured from \galex\ FUV and NUV photometry.

Our sample consisted of the 28 stars which have both reconstructed \lya\ and \galex\ observations ranging in SpT from F5 to M5.5. We separated our sample into 11 F and G stars and 17 K and M stars as the photospheric contribution to the \galex\ bandpasses, stellar variability, and chromospheric temperature structure (e.g.~\citealt{walk09}) differ in the two groups. The F and G stars were limited to only eight FUV observations as they were too bright for reliable NUV measurements. There is a weak correlation between the FUV excess flux and \lya, and a strong one between \galex/FUV and the Mg II  doublet, the strongest emission feature in the NUV spectral region. However, with the few data points for F and G stars, their small flux distribution, and uncertainties in the model fluxes, these correlations should be applied cautiously. 

For the sample of K and M stars where the photospheric model fluxes are negligible, we find that the \lya\ surface fluxes correlate well with both the FUV and NUV excess surface fluxes such that  log [$F_{Ly\alpha}$] = (0.43 $\pm$ 0.07) * log [$F_{FUV, exc}$] + (3.97 $\pm$ 0.36)	and log [$F_{Ly\alpha}$] = (0.45 $\pm$ 0.07) * log [$F_{NUV, exc}$] + (3.55 $\pm$ 0.41). For stars too bright in the NUV, the FUV correlation can be used, and for more distant stars where the FUV is not detected, the NUV correlation can be used. Additionally, the \galex\ excess fluxes for K and M stars correlate well with the Mg II lines providing another window into a strong stellar chromospheric emission feature probing lower emission temperatures.  

With good correlations in both \galex\ bandpasses, estimates of intrinsic \lya\ and Mg II can now be made for thousands of K and M stars out to a few hundred parsecs from Earth. These data will constrain a new and much-needed suite of model chromospheres, as well as provide planetary atmosphere models with more realistic \lya\ and Mg II values without the need for high-resolution UV spectroscopy.

 

%%%%%%%%%%%%%%%%%%%%%%%%%%%%%%%%%%%%%%%%%%%%%%%%%%%%%%%%%%%%%%%%%%%%%%%%%%%%%%%%%%%%%%%%%%%%%%%%%%%%%%%%%%%%%%%%%%%%%%%%%%%%%%
\acknowledgements

We thank B.~Skiff for helpful comments and K.R.~acknowledges funds from the US NSF Research Experience for Undergraduates at Northern Arizona University. We also thank the anonymous referee for her/his comments. The research conducted made use of the VizieR catalogue, SIMBAD database, the Mikulski Archive for Space Telescopes (MAST), and the Phoenix and Kurucz photosphere-only atmospheric models.  This research has made use of the VizieR catalogue access tool, CDS, Strasbourg, France \citep{ochs00} and the Mikulski Archive for Space Telescopes (MAST). STScI is operated by the Association of Universities for Research in Astronomy, Inc., under NASA contract NAS5-26555. Support for MAST for non-HST data is provided by the NASA Office of Space Science via grant NNX13AC07G and by other grants and contracts.

%%%%%%%%%%FIGURES%%%%%%%%%%%%%%%%%%

\clearpage        
	
%%%%%%%%%%%%%%%%%%%%%%%%%%%%%%%%%%%%%%%%%%%%%%%%%

%\bibliography{/Users/evgenyashkolnik/Dropbox/refs_master}{}
%\bibliographystyle{apj}

\clearpage

\begin{deluxetable}{lllccccccccccc}																																																							
\tabletypesize{\scriptsize}																																																							
%\rotate																																																							
\tablecaption{Stellar Properties and Surface Line Fluxes  \label{table_targets}}																																																							
\tablewidth{0pt}																																																							
\tablehead{																		
\colhead{Name} 	&	 \colhead{SpT\tablenotemark{a}} 	&	\colhead{Dist.\tablenotemark{a}} 	&	 \colhead{$R_{*}$\tablenotemark{b}} 	&	 \colhead{\Teff\tablenotemark{c}}	&	 \colhead{log($F_{NUV, obs}$)}	&	 \colhead{log($F_{NUV, exc}$)}	&	 \colhead{log($F_{FUV, obs}$)}	&	 \colhead{log($F_{FUV, exc}$)}	\\
\colhead{} 		&	 \colhead{} 	&	\colhead{pc}	&	 \colhead{R$_{\odot}$} 	&	 \colhead{K} 	&	 \colhead{ erg cm$^{-2}$ s$^{-1}$} 	&	 \colhead{ erg cm$^{-2}$ s$^{-1}$} 	&	 \colhead{ erg cm$^{-2}$ s$^{-1}$} 	&	 \colhead{ erg cm$^{-2}$ s$^{-1}$} 	
}																																																		
\startdata																		
HR 4657	&	F5 V/L	&	22.6	&	1.4	&	6440	&	too bright	&	--	&	6.62 $\pm$  0.06	&	$<$ 6.05	\\
Chi Ori	&	G0 V	&	8.7	&	1.3	&	6030	&	too bright	&	--	&	not observed	&	--	\\
HR 4345	&	G0 V	&	21.9	&	1.3	&	6030	&	too bright	&	--	&	5.77 $\pm$  0.06	&	5.67 $\pm$  0.09	\\
V376 Peg	&	G0 V	&	49.6	&	1.4	&	6030	&	too bright	&	--	&	5.74 $\pm$  0.07	&	5.65 $\pm$  0.11	\\
HR 2882	&	G4 V	&	21.8	&	1.2	&	5835	&	too bright	&	--	&	not observed	&	--	\\
61 Vir	&	G5 V	&	8.6	&	1.4	&	5770	&	too bright	&	--	&	4.85 $\pm$  0.07	&	4.74 $\pm$  0.11	\\
HR 2225	&	G5 V	&	16.7	&	1.2	&	5770	&	too bright	&	--	&	5.51 $\pm$  0.06	&	5.48 $\pm$  0.07	\\
HD 203244	&	G5 V	&	20.4	&	1.2	&	5770	&	too bright	&	--	&	5.38 $\pm$  0.05	&	5.34 $\pm$  0.06	\\
HD 128987	&	G6 V	&	23.7	&	1.1	&	5670	&	too bright	&	--	&	5.32 $\pm$  0.05	&	5.29 $\pm$  0.06	\\
Xi Boo A	&	G8 V	&	6.7	&	1.1	&	5570	&	too bright	&	--	&	5.54 $\pm$  0.03	&	5.53 $\pm$  0.04	\\
HD 116956	&	G9 IV-V	&	21.9	&	1.0	&	5410	&	too bright	&	--	&	5.48 $\pm$  0.06	&	5.48 $\pm$  0.07	\\
DX Leo	&	K0 V	&	17.8	&	0.9	&	5250	&	too bright	&	--	&	5.70 $\pm$  0.04	&	5.70 $\pm$  0.04	\\
HR 8	&	K0 V	&	13.7	&	1.1	&	5250	&	too bright	&	--	&	not observed	&	--	\\
Epsilon Eri	&	K1 V	&	3.2	&	0.9	&	5075	&	too bright	&	--	&	5.11 $\pm$  0.03	&	5.11 $\pm$  0.03	\\
40 Eri A	&	K1 V	&	5.0	&	1.1	&	5075	&	too bright	&	--	&	too bright	&	--	\\
HR 1925	&	K1 V	&	12.3	&	0.9	&	5075	&	too bright	&	--	&	5.21 $\pm$  0.05	&	5.21 $\pm$  0.05	\\
EP Eri	&	K2 V	&	10.4	&	0.9	&	4900	&	too bright	&	--	&	5.45 $\pm$  0.04	&	5.45 $\pm$  0.04	\\
LQ Hya	&	K2 V	&	18.6	&	0.9	&	4900	&	too bright	&	--	&	6.18 $\pm$  0.04	&	6.18 $\pm$  0.04	\\
V368 Cep	&	K2 V	&	19.2	&	1.1	&	4900	&	too bright	&	--	&	not observed	&	--	\\
PW And	&	K2 V	&	21.9	&	1.1	&	4900	&	7.05 $\pm$  0.07	&	6.64 $\pm$  0.19	&	not observed	&	--	\\
Speedy Mic	&	K2 V/L	&	52.2	&	1.1	&	4900	&	7.72 $\pm$  0.22	&	7.66 $\pm$  0.26	&	6.37 $\pm$  0.22	&	6.37 $\pm$  0.22	\\
Epsilon Ind	&	K5 V	&	3.6	&	0.8	&	4350	&	too bright	&	--	&	4.63 $\pm$  0.06	&	4.63 $\pm$  0.06	\\
AU Mic	&	M0 V	&	9.9	&	1.0	&	3850	&	6.30 $\pm$  0.06	&	6.29 $\pm$  0.06	&	5.48 $\pm$  0.06	&	5.48 $\pm$  0.06	\\
GJ 832	&	M1.5 V	&	5.0	&	0.4	&	3680	&	5.28 $\pm$  0.06	&	5.13 $\pm$  0.09	&	4.26 $\pm$  0.09	&	4.26 $\pm$  0.09	\\
GJ 436	&	M3 V	&	10.3	&	0.2	&	3350	&	5.47 $\pm$  0.11	&	5.43 $\pm$  0.13	&	$<$ 4.97	&	$<$ 4.97	\\
AD Leo	&	M3.5 V	&	4.7	&	0.3	&	3265	&	6.54 $\pm$  0.14	&	6.54 $\pm$  0.14	&	not observed	&	--	\\
GJ 876	&	M5.0 V	&	4.7	&	0.3	&	3010	&	4.77 $\pm$  0.21	&	4.38 $\pm$  0.68	&	3.37 $\pm$  0.37	&	3.37 $\pm$  0.37	\\
Proxima Cen	&	M5.5 V	&	1.3	&	0.2	&	2925	&	4.93 $\pm$  0.17	&	4.90 $\pm$  0.18	&	not observed	&	--	
																	
\enddata																																																										
\tablenotetext{a}{Data compiled by \cite{lins13}.}																																																																																																														
\tablenotetext{b}{Stellar radii determined from \cite{bara98} models using stellar ages and \Teff.}																																																																																																														
\tablenotetext{c}{\Teff\ from \cite{krau07}.}																	
\end{deluxetable}

	\begin{deluxetable}{lllccccccccccc}																																																
	\tabletypesize{\scriptsize}																																																
	%\rotate																																																
	\tablecaption{Regression Coefficients  \label{table_coef}}																																																
	\tablewidth{0pt}																																																
	\tablehead{											
	\colhead{Regression Fit\tablenotemark{a}} 	&	 \colhead{SpT} 	&	 \colhead{\#} 	&	 \colhead{A}	&	 \colhead{B} 	&	 \colhead{$R$}  \\
	\colhead{} 		&	 \colhead{Range} 	&	 \colhead{} 	&	 \colhead{} 	&	 \colhead{} 	&	 \colhead{} 
	}																																											
	\startdata																				
	log [$F_{NUV, phot}$] = A \Teff\ + B	&	F5 V - M5.5 V	&	28	&	1.519E--03 $\pm$ 4.080E--05	&	--0.6094 $\pm$ 0.2063	&	0.99 \\
	log [$F_{FUV, phot}$] = A \Teff\ + B	&		F5 V - M5.5 V	&	28	&	2.640E--03 $\pm$ 6.484E--05	&	--10.6934 $\pm$ 0.3279	&	0.99 \\
	\hline	&		&		&		&		& \\
	 log [$F_{Ly\alpha}$] = A log [$F_{NUV, obs}$] + B	&	K2 V - M5.5 V	&	8	&	0.430 $\pm$ 0.094	&	3.595 $\pm$ 0.575	&	0.88 \\
	log [$F_{Ly\alpha}$] = A log [$F_{NUV, exc}$] + B	&	K2 V - M5.5 V	&	8	&	0.447 $\pm$ 0.069	&	3.549 $\pm$ 0.409	&	0.94 \\
	 log [$F_{Ly\alpha}$] = A log [$F_{FUV, exc}$] + B	&	K0 V - M5.5 V	&	10	&	0.432 $\pm$ 0.068	&	3.969 $\pm$ 0.359	&	0.91 \\
	 log [$F_{Ly\alpha}$] = A log [$F_{FUV, exc}$] + B	&	F5 V - G9 V	&	8	&	0.853 $\pm$ 0.427	&	1.325 $\pm$ 2.307	&	0.63 \\
	\hline	&		&		&		&		& 	\\
	 log [$F_{Ly\alpha}$/$F_{NUV, exc}$] = A log [$F_{NUV, exc}$] + B	&	K2 V - M5.5 V	&	8	&	--0.556 $\pm$ 0.069	&	3.570 $\pm$ 0.411	&	-0.96 \\
	 log [$F_{Ly\alpha}$/$F_{FUV, exc}$] = A log [$F_{FUV, exc}$] + B	&	K0 V - M5.5 V	&	10	&	--0.570 $\pm$ 0.071	&	3.970 $\pm$ 0.374	&	-0.94 \\
	 log [$F_{Ly\alpha}$/$F_{FUV, exc}$] = A log [$F_{FUV, exc}$] + B	&	F5 V - G9 V	&	8	&	no corr. 	&	--	&	-- \\
	\hline	&		&		&		&		& 	\\
	 log [$F_{MgII}$] = A log [$F_{NUV, exc}$] + B	&	K2 V - M5.5 V	&	7	&	0.872 $\pm$ 0.049	&	0.349 $\pm$ 0.301	&	0.99 \\
	 log [$F_{MgII}$] = A log [$F_{FUV, exc}$] + B	&	K0 V - M5.5 V	&	10	&	0.916 $\pm$ 0.116	&	1.272 $\pm$ 0.610	&	0.94 \\
	 log [$F_{MgII}$] = A log [$F_{FUV, exc}$] + B &	F5 V - G9 V	&	6	&	1.056 $\pm$ 0.177	&	0.599 $\pm$ 0.946	&	0.95 \\
	\enddata																																																			
	\tablenotetext{a}{Surface flux units are in ergs cm$^{-2}$ s$^{-1}$.}																																																																																																							
%	\tablenotetext{b}{}																																																																																																							
%	\tablenotetext{d}{}										
	\end{deluxetable}

\end{document}